\documentclass{moriond}
\usepackage{amsmath}
\usepackage{amssymb}
\usepackage{color}
\usepackage{wrapfig}
\usepackage{subfigure}
\usepackage{cite}

\def\Journal#1#2#3#4{{#1} {\bf #2}, #3 (#4)}

\def\JHEP{\em JHEP}

\def\PLB{{\em Phys. Lett.}  B}
\def\PRL{\em Phys. Rev. Lett.}
\def\PRD{{\em Phys. Rev.} D}

\def\EPJC{{\em Eur. Phys. J.} C}

\def\be{\begin{equation}}
\def\ee{\end{equation}}
\def\bea{\begin{eqnarray}}
\def\eea{\end{eqnarray}}

\def\bpm{\begin{pmatrix}}
\def\epm{\end{pmatrix}}

\newcommand{\fig}[1]{Fig.~\ref{#1}}
\newcommand{\eq}[1]{Eq.~\ref{#1}}
\newcommand{\tab}[1]{Tab.~\ref{#1}}
 
\newcommand\lr{\left( } \newcommand\rr{\right) }

\newcommand\lc{\left\{ } \newcommand\rc{\right\} }
\newcommand{\gau}[1][]{\ensuremath{\widetilde{\chi}^{#1}}} 
\newcommand{\neu}[1][]{\ensuremath{\gau[0]_{#1}}}	
\newcommand{\cha}[1][]{\ensuremath{\gau[\pm]_{#1}}} 

\newcommand{\Photo}{}

\begin{document}
\vspace*{4cm}
\title{SIMPLIFIED GAUGINO-HIGGSINO MODELS IN THE MSSM}

\author{ M.P.A. SUNDER }

\address{Institut f\"{u}r Theoretische Physik, Westf\"{a}lische Wilhelms-Universit\"{a}t M\"{u}nster,\\ 
Wilhelm-Klemm-Stra{\ss}e 9, 48149 M\"{u}nster, Germany}

\maketitle\abstracts{We present a tool to produce benchmarks with realistic mass spectra and 
realistic mixing in the gaugino-higgsino sector of the MSSM. We suggest as a next-to-minimal 
approach the use of benchmarks, whose mass spectra and mixing matrix elements are the result 
of a proper matrix diagonalisation at treelevel. We scan over the four relevant parameters 
$\{\mu, \tan \beta, M_{1}, M_{2}\}$ for a specific grid of neutralino and chargino masses. We 
demonstrate how to define a measure for the quality of a fit, including a method to maximise
properties such as the gaugino or higgsino content.}

\section{The gaugino-higgsino sector}

The gauginos and higgsinos are the fermionic superpartners of the electroweak gauge-bosons and the
two $SU\lr 2\rr$-valued complex scalar Higgs doublets that appear in the Minimal Supersymmetric 
Standard Model (MSSM). Through electroweak symmetry breaking (EWSB) these fermionic states mix
to form either neutral neutralinos \neu[i] \ or electrically charged charginos \cha[i]. 

\noindent The neutralinos are mixed states of neutral bino $\widetilde{B}^{0}$, wino $\widetilde{W}_{3}^{0}$ and 
higgsino fields $\widetilde{H}^{0}_{i}$,

\be 
\bpm \widetilde{B}^{0} & \widetilde{W}_{3}^{0} & \widetilde{H}^{0}_{1} & \widetilde{H}^{0}_{2} \epm^{\text{T}}  \xrightarrow{\text{EWSB}} \bpm \neu[1] & \neu[2] & \neu[3] & \neu[4] \epm^{\text{T}}  \ \text{.}
\ee

\noindent The charginos are mixed states of charged winos $\widetilde{W}_{i}^{\pm}$ and higgsino fields 
$\widetilde{H}^{\pm}_i$,

\be
\lc \bpm \widetilde{W}^{-} \\ \widetilde{H}^{-}_{1} \epm \text{,}  \bpm \widetilde{W}^{+} \\ \widetilde{H}^{+}_{2} \epm \rc  \xrightarrow{\text{EWSB}} \lc \bpm \chi^{-}_{1} \\ \chi^{-}_{2} \epm \text{,}  \bpm \chi^{+}_{1} \\ \chi^{+}_{2} \epm \rc \ \text{.}
\ee 

\noindent The other particles in the MSSM, such as the gluino and the squarks, are decoupled at values of
around $1\text{-}1.5~\text{TeV}$, in accordance with the current mass limits from supersymmetry (SUSY) searches~\cite{atlas170904,atlas170809,atlas170808,cms170703,cms170707,cms170604}. We shall
focus on simplified MSSM models that have neutralinos and charginos as their lightest supersymmetric 
particles, considered in experimental studies such as~\cite{cms170908,atlas170807,cms170904}. 

These simplified gaugino-higgsino models are governed by only four additional MSSM parameters,
which gives them some predictive value. These parameters are,

\be \label{eq:def_params}
\{ \overbrace{\mu}^{\mathcal{W}_{\text{MSSM}}}, \overbrace{\tan \beta}^{\text{EWSB}}, 
\overbrace{M_{1}, M_{2}}^{\text{SSB}} \} \ \text{,}
\ee

\noindent where $\mu$ is the effective higgsino mass parameter originating from the superpotential
$\mathcal{W}_{\text{MSSM}}$, $\tan \beta$ is the ratio of the two scalar Higgs vacuum expectation values
(vevs) $H^{0}_2$ and $H^{0}_1$, whilst $M_1$ and $M_2$ are the gaugino mass parameters 
that originate from the soft supersymmetry breaking (SSB).

\section{The minimal vs. non-minimal approach}

The minimal approach for SUSY searches in this sector, e.g. as used by
~\cite{atlas17orb,cms131233,cms151208,cms170904}, is to set a mass spectrum without any actual 
mixing among the gauginos and between gauginos and higgsinos. This approach is applicable in 
three different scenarios, namely in case of: 

\begin{wrapfigure}{r}{5.0cm}
  \includegraphics[width=0.25\textwidth]{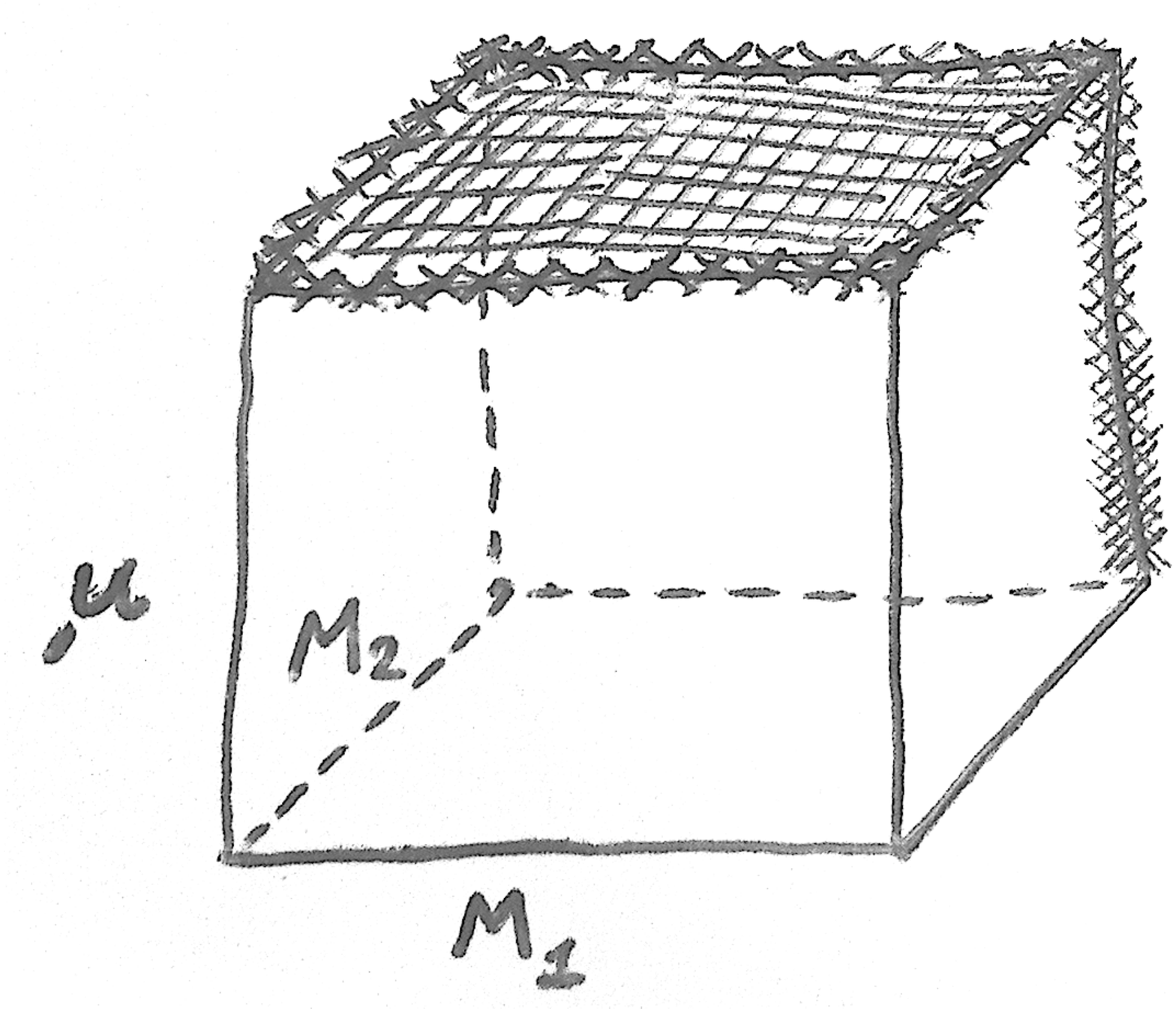}
  \caption{Here, we visualise the regions of parameter space wherein the minimal approach is valid, 
  these are the shaded areas corresponding to either $\mu \rightarrow \infty$ or both $\lc M_1, M_2 
  \rc \rightarrow \infty$.} \label{fig:param_rngs}
\end{wrapfigure}

\begin{itemize} \itemsep-0.5em
\item a bino lightest supersymmetric particle (LSP) with mass $M_{\neu[1]} = M_1$ and an additional 
doublet of degenerate wino next-to lightest supersymmetric particles (nLSPs) at $\lc M_{\neu[2]}, 
M_{\cha[1]}\rc= M_2$, which requires $M_1 < M_2$ and $\mu \rightarrow \infty$.
\item a doublet of degenerate wino LSPs with masses $\lc M_{\neu[1]}, M_{\cha[1]}\rc= M_2$ and an 
additional bino nLSP at $M_{\neu[2]} = M_1$, which requires $M_2 < M_1$ and $\mu \rightarrow \infty$.
\item a triplet of degenerate higgsino LSPs with masses $\lc M_{\neu[1]}, M_{\neu[2]}, M_{\cha[1]}\rc= 
\mu$, which requires that $\lc M_1, M_2 \rc \rightarrow \infty$.
\end{itemize}

Independently of $\tan \beta$ these three scenarios are applicable in two regions of the  
parameter space spanned by $\lc \mu, M_1, M_2 \rc$ as depicted in \fig{fig:param_rngs}. We 
suggest a next-to-minimal approach in~\cite{fuks171009} to obtain benchmarks that are applicable 
in cases where gaugino and higgsino states mix, as demonstrated in~\cite{cms180101}. 
This method is applicable in the whole parameter space depicted in \fig{fig:param_rngs}.

Instead of setting the mass spectrum, we scan over the parameters in \eq{eq:def_params}
and compute the mass spectrum by proper diagonalisation of the mass matrices at tree-level. 
In this way, benchmarks that are extremely fine-tuned or unphysical will not be found, one 
can explore the relationships between mass-splittings and coupling strengths by using the 
appropriate mixing matrix elements and constrain certain regions in the MSSM parameter 
space more directly.

The search region must be defined, whilst taking into account approximate relations and 
parameter transformations that do not affect the mass spectra or particle content.  

\section{Case-Study: \textit{Higgsino-like benchmarks with equidistant mass splitting}}

We present an example scenario of higgsino-like benchmarks with equidistant mass splitting
to clarify how one sets up a parameter scan for a specific scenario. 

The chosen scenario of an equidistant mass-splitting between the chargino $\cha[1]$ and the 
two neutralinos $\neu[1]$ and $\neu[2]$ motivates a parametrisation of the mass spectrum in
terms of $M_{\cha[1]}$, the mass of $\cha[1]$ at the intermediate scale, and $\Delta M_{21}$, 
the total size of the mass-splitting between $\neu[1]$ and $\neu[2]$. The neutralino masses
are then given by,

\be 
M_{\neu[1]}=M_{\cha[1]}-\frac{\Delta M_{21}}{2} \quad \text{and} \quad M_{\neu[2]}=M_{\cha[1]}
     +\frac{\Delta M_{21}}{2} \ \text{.}
\ee

This parametrisation was used to define a grid from which we sampled $\mathcal{O}\lr 10^4 \rr$ 
benchmarks in a region that would be accessible for LHC searches given in \tab{tab:massgrid}. 
The targeted spectra in \tab{tab:massgrid} and the general parameter dependencies of the mass
spectrum motivate an initial scan-range given by, 

\be\begin{matrix}
|\mu| \in [ \text{min}(M_{\cha[1]})-\mathcal{O}(M_{W}),\text{max}(M_{\cha[1]})+\mathcal{O}(M_{W})] \ \text{,} \\ 
\{|M_1|,M_2\} \in [ \text{min}(M_{\cha[1]})-\mathcal{O}(M_{W}),5~\text{TeV}] \ \text{,} \ \tan\beta \in  [0,100] \ \text{,}
\end{matrix}\ee

\noindent where either sign of the $\mu$-parameter yields equally good benchmarks, though with
a lower higgsino content once $\Delta M_{21}\gtrsim 25~\text{GeV}$.

The randomly selected benchmarks were scored using three dimensionless selection criteria, as 
shown in \eq{eq:selcrit}. The deviation between found and targeted benchmarks was constrained 
by an upper limit on the benchmark's score at $\text{score}_{\text{max}} = 0.1$. 
 
\begin{table}
 \caption{\label{tab:massgrid}The targeted chargino masses and mass-splitting in the parameter scan 
 for the scenario of higgsino-like benchmarks with equidistant mass-splittings. The $\Delta\lr \ldots \rr$
 column refers to the used grid-spacing of either $M_{\cha[1]}$ or $\Delta M_{21}$.}
 \begin{center} \begin{tabular}{|c|rrr|}
   \hline
    Targeted mass(-splitting)  & Min.  & Max.  & $\Delta\lr \ldots \rr$ \\
    \hline
    $M_{\cha[1]}$ & $90$ GeV & $400$ GeV & $3.1$ GeV \\ 
    $\Delta M_{21}$         & $1$ GeV   & $100$ GeV & 1 GeV\\
    \hline
  \end{tabular} \end{center} \vspace*{-6mm} \end{table}

\be \label{eq:selcrit}
\left. \begin{matrix} d_1 \Delta M_{\chi^{\pm}_{1}} =& \delta M_{\chi^{\pm}_{1}} \\
 d_2 \Delta(\Delta M_{21}) =& 2\delta (M_{\chi^{0}_{2}}-M_{\chi^{\pm}_{1}}) \\ 
 d_3 \Delta(\Delta M_{21}) =& 2\delta (M_{\chi^{\pm}_{1}}-M_{\chi^{0}_{1}}) \end{matrix} \rc \ 
 \Rightarrow \text{score} = \left. \sqrt{\frac{d_{1}^{2}+d^{2}_{2}+d^{2}_{3}}{3}} \right|_{\text{acceptable}} < 0.1
\ee

The average higgsino content $\widetilde{f}$ of $\lc \neu[1],\cha[1],\neu[2]\rc$ was maximised by
reweighting the score with 

\be
\text{score}_{\text{new}} \frac{\widetilde{f}_{\text{old}}}{\widetilde{f}_{\text{new}}} < \text{score}_{\text{old}} \ \text{,}
\ee

\noindent which preferentially selects the benchmark with the higher higgsino content in case of a 
comparable agreement with the targeted mass spectrum. A redefinition of $\widetilde{f}$ is usable 
to maximise other benchmark properties.   

\begin{figure}[!h]\begin{center}
    \subfigure{\includegraphics[width=0.4\textwidth]{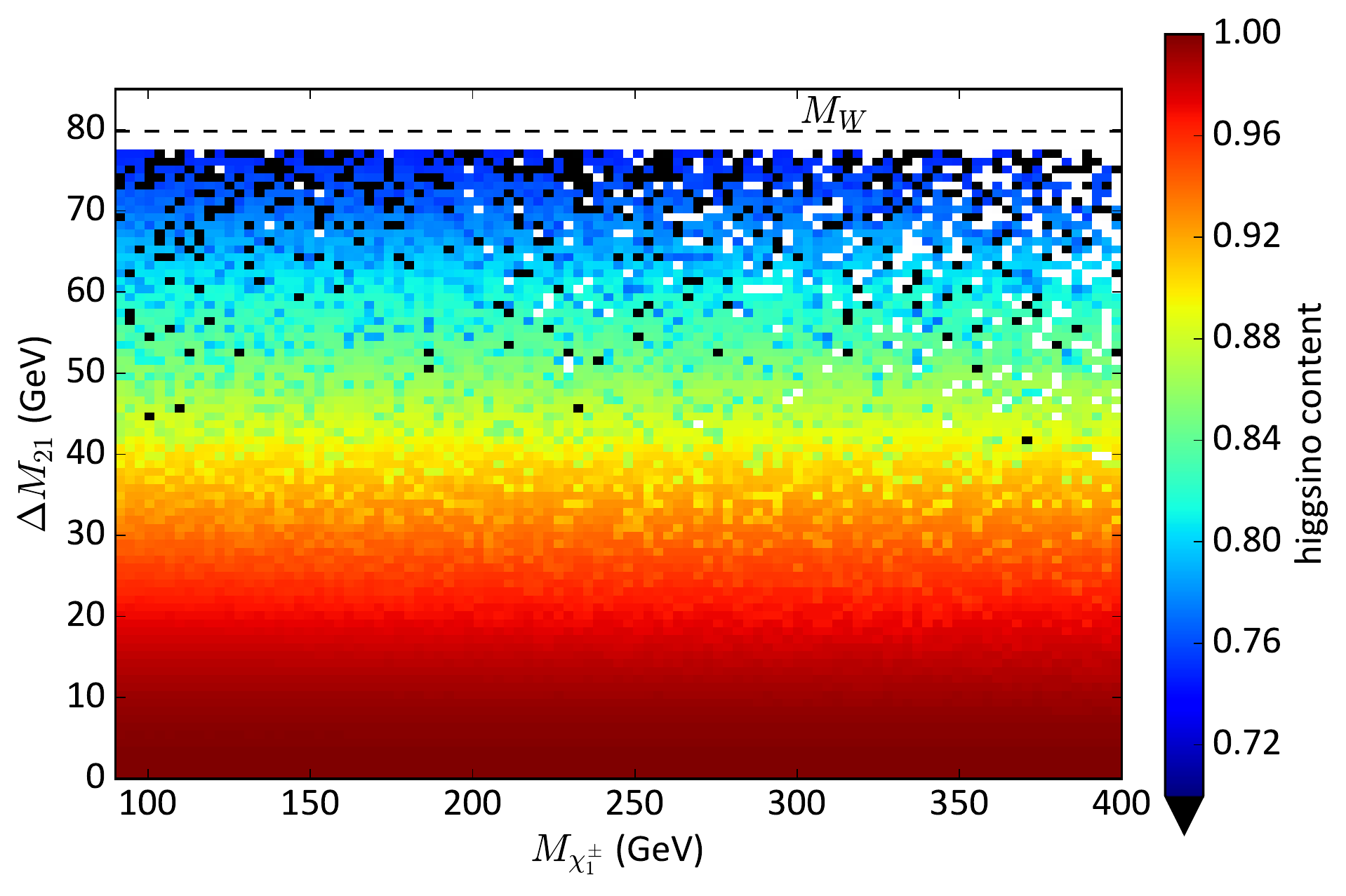}}
     \subfigure{\includegraphics[width=0.4\textwidth]{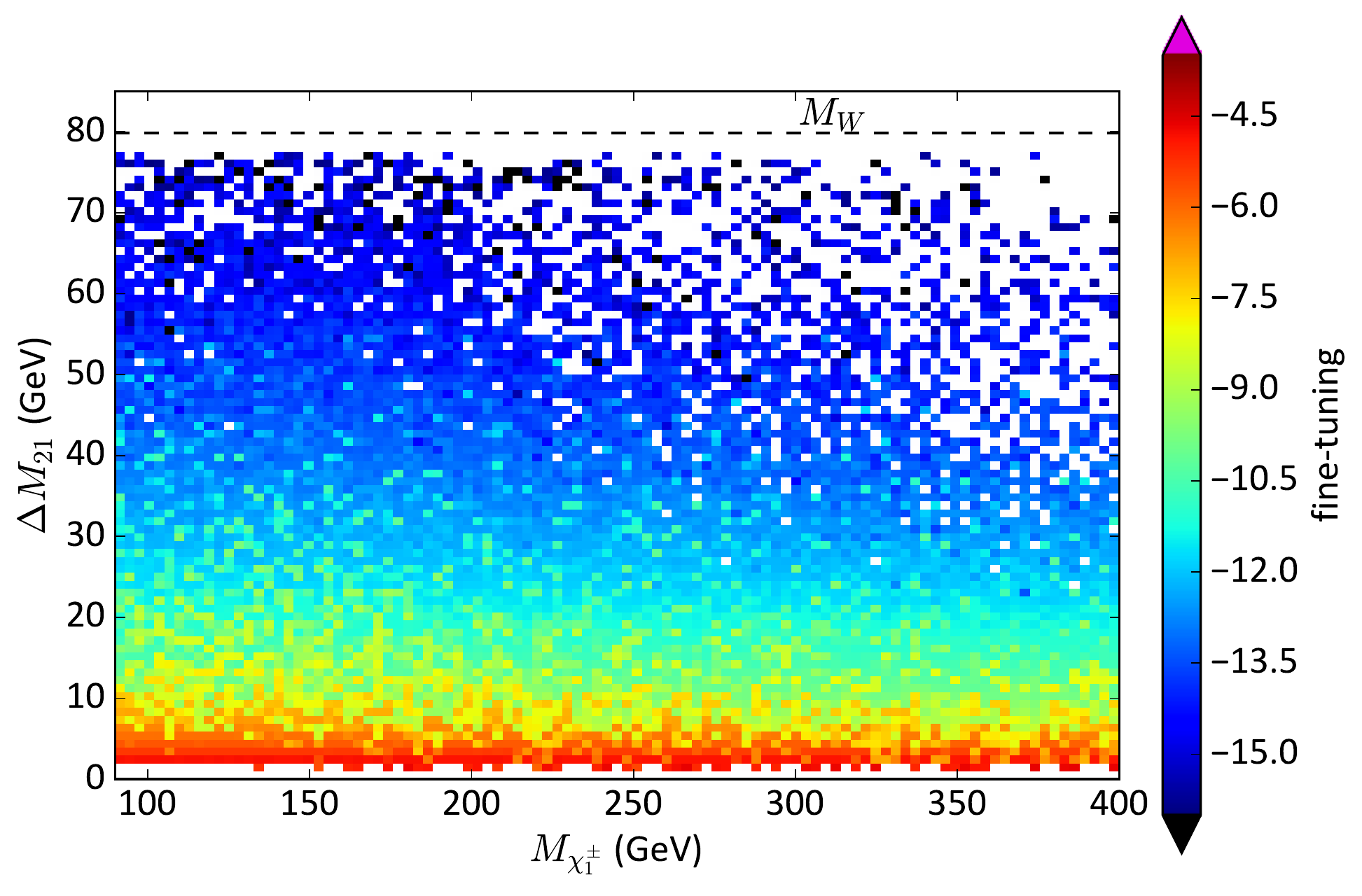}}
\caption{This figure shows the average higgsino content of $\lc \neu[1],\cha[1],\neu[2]\rc$ of and a 
fine-tuning measure of the benchmarks that were found for the subscenario wherein $\mu>0$. 
If the fine-tuning $\widetilde{f}$ for two benchmarks relates as $\widetilde{f}\lr b_2 \rr - \widetilde{f}
\lr b_1 \rr = 5$ then the number of found acceptable benchmarks for $b_2$ is a factor $10^5$ less 
than for $b_1$.}\label{fig:scanresults}     
\end{center}\end{figure}

In \fig{fig:scanresults} we demonstrate the feasibility of finding benchmarks for the subscenario with $\mu>0$.
We succeeded in finding benchmarks with reasonable higgsino contents of $\gtrsim 0.7$ for $\Delta M_{21} 
\lesssim M_{W}$ independent of the targeted $M_{\cha[1]}$. Although, from $\Delta M_{21} \gtrsim 40~\text{GeV}$
the amount of fine-tuning made it increasingly more difficult to find these benchmarks, which can be seen by 
the absence of benchmarks or the poor maximisation of the higgsino content. The fine-tuning measure was chosen
as the logarithms of the product of the allowed acceptable variation of the benchmark parameters divided by the total
search range for each of those parameters. 

\section{Conclusions}

We argue that the discussed minimal approach has only a limited applicability and does not constrain directly
the parameter space of the MSSM in the gaugino-higgsino sector. We suggested here and in~\cite{fuks171009} 
a next-to-minimal approach wherein the whole parameter space in the gaugino-higgsino sector can be explored 
by scanning over the MSSM parameters $\lc \mu, \tan\beta, M_1, M_2 \rc$. Use of this approach constrains the 
MSSM parameter space more directly and guarantees that the used benchmark is representative of a true 
non-simplified MSSM benchmark. We demonstrate that this approach is feasible in finding a high resolution grid of 
benchmarks for the particular scenario of higgsino-like benchmarks with equidistant mass-splitting, which 
would not be treatable in the minimal approach.

\section*{Acknowledgments}

This work has been supported by the BMBF under contract 05H15PMCCA and the DFG through the 
Research Training Network 2149 "Strong and weak interactions - from hadrons to dark matter".

\end{document}